\documentclass[%
reprint,
superscriptaddress,
amsmath,amssymb,
aps,
pra,
floatfix,
]{revtex4-2}

\usepackage{xcolor}
\usepackage{lipsum}
\usepackage{braket}
\usepackage{graphicx}
\usepackage{dcolumn}
\usepackage{bm}

\newcommand{\imag}{\mathrm{i}}

\begin{document}

\title{No need for a grid: Adaptive fully-flexible gaussians for solving the time-dependent Schrödinger equation}

\author{Simen Kvaal}
\email{simen.kvaal@kjemi.uio.no}
\affiliation{Centre for Advanced Study at the Norwegian Academy of Science and Letters, Drammensveien 78, N-0271 Oslo,
Norway}
\affiliation{Hylleraas Centre for Quantum Molecular Sciences, Department of Chemistry, University of Oslo, Norway}
\author{Caroline Lasser}
\affiliation{Centre for Advanced Study at the Norwegian Academy of Science and Letters, Drammensveien 78, N-0271 Oslo,
Norway}
\affiliation{Zentrum Mathematik, Technische Universität München, München, Germany}
\author{Thomas Bondo Pedersen}
\affiliation{Centre for Advanced Study at the Norwegian Academy of Science and Letters, Drammensveien 78, N-0271 Oslo,
Norway}
\affiliation{Hylleraas Centre for Quantum Molecular Sciences, Department of Chemistry, University of Oslo, Norway}
\author{Ludwik Adamowicz}
\affiliation{Centre for Advanced Study at the Norwegian Academy of Science and Letters, Drammensveien 78, N-0271 Oslo,
Norway}
\affiliation{Department of Chemistry and Biochemistry, University of Arizona, Tucson, Arizona 85721, USA}

\date{7 March 2023}

\begin{abstract}
Linear combinations of complex gaussian functions, where the linear and nonlinear parameters are allowed to vary, are shown to 
provide an extremely flexible and effective approach for solving the time-dependent Schrödinger equation in one spatial dimension.
The use of flexible basis sets has been proven notoriously hard within the systematics of the Dirac--Frenkel variational principle.
In this work we present an alternative time-propagation scheme that de-emphasizes optimal parameter evolution but directly targets residual minimization via the method of Rothe's method, also called the method of vertical time layers. We test the scheme using a simple model system mimicking an atom subjected to an extreme laser pulse. Such a pulse produces
complicated ionization dynamics of the system.  The scheme is shown to perform very well on this model and notably does not rely
on a computational grid.
Only a handful of gaussian functions are needed to achieve an accuracy on par with a high-resolution, grid-based solver.
This paves the way for accurate and affordable solution of the time-dependent Schrödinger equation for atoms and molecules
within and beyond the Born--Oppenheimer approximation.
\end{abstract}

\maketitle


\newcommand{\note}[1]{{\color{red}#1}}
\newcommand{\argmin}{\operatornamewithlimits{argmin}}

\section{Introduction}
\label{sec:introduction}

When atoms and molecules are subjected to ultrashort and intense laser pulses, their wave functions become highly complicated and fast-changing
due to ionization and fragmentation processes~\cite{Nisoli2017_CR}.
High spatial and temporal resolutions and large computational domains are mandatory when the time-dependent Schrödinger equation (TDSE)
is solved numerically, and computational approaches that, at some level, are based on the Born--Oppenheimer approximation may no
longer be appropriate.
For even the most modestly sized atoms, it is simply impossible
to achieve experiment-level accuracy with a grid or an otherwise fixed basis due to the curse of dimensionality of the problem.
Therefore, it is essential to develop compact and flexible wave-function representations that can be efficiently propagated
in a numerically stable manner.

This work has two main aims: The first is to demonstrate that 
gaussian functions with time-evolving complex parameters form a very flexible and efficient basis set
to represent complicated wave-function dynamics. The
second is to present a computational method for optimizing the nonlinear parameters of the gaussians whilst changing 
their number as needed during the simulation.
As an initial test, we consider a 1D model 
problem mimicking a hydrogenic atom or ion subjected to an ultrashort laser pulse that
induces rather extreme dynamics, including ionization.
With the new time-propagation scheme based on nonlinear 
least-squares optimization, we show that a compact gaussian basis set can reproduce 
the essentially exact grid-based reference
solution of the TDSE for our model problem.
In particular, the ionization tail of the wave function is very well 
captured at every stage of the dynamics using only a handful of the complex 
gaussians.

Using a linear combination of gaussians (LCG) \emph{Ansatz}
with complex and in particular explicitly correlated parameters 
is well established for  
highly accurate calculations of bound states of small atoms and molecules 
using the Rayleigh--Ritz variational method~\cite{Mitroy2013_RMP}.
Notably, some of these applications have been performed \emph{without} assuming
the Born--Oppenheimer approximation, i.e., treating the motion of the nuclei and 
electrons on an equal footing.
An attractive feature of the LCG \emph{Ansatz} is the existence of analytic formulas for matrix elements involving complex 
gaussians, including the explicitly correlated ones~\cite{Bubin2006_TJoCP}. 

The time-dependent Dirac--Frenkel principle \cite{Dirac1930,Frenkel1934,McLachlan1964,Kramer1981} appears as
a natural framework for a dynamical extension of the variational LCG approach to bound states.
The Dirac--Frenkel principle, however, has a notorious problem 
with severely ill-conditioned or even singular Gramian matrices \cite{Kay1989}.
This problem is particularly acute for LCG wave functions~\cite{Rowan2020_PRE} and yields to 
the frozen width of the gaussians and an additional Thykonov regularization in the variational
multi-configurational gaussian (vMCG)~\cite{RPSWBL2015} method.  
Regrettably, the great flexibility of the complex gaussians comes at a prize.

The alternative time-propagation scheme proposed in this work averts the matrix singularity 
problem of the Dirac--Frenkel principle by using Rothe's method, also called the method of vertical time layers~\cite{Rothe1930_MA,Deuflhard2012_},
for solving the TDSE.
The Rothe method is a variational approach that, in a natural manner, allows 
for adaptivity of the size of the gaussian basis set during the
time-propagation of the wave function: The number of gaussians
is increased whenever the error in the wave function becomes too large. 
Moreover, the method is entirely formulated in terms of the gaussian matrix elements of 
certain operators and their derivatives with respect to the nonlinear parameters of the gaussians. 
Thus, the method \emph{completely eliminates} the need for a grid, while 
allowing the wave function to freely roam space with, in principle, infinite level of detail.

The compression levels observed with LCG wave functions 
for the 1D case in the present work may be even 
more pronounced for multi-particle systems 
with the use of $N$-particle explicitly correlated 
gaussians that have been routinely employed 
in calculating bound spectra of small atoms and molecules with very high accuracy. 
The method presented here thus paves the way for highly accurate 
solutions of the TDSE for realistic chemical systems, including ionization and
dissociation processes induced by sub-femtosecond laser pulses,
employing neither a grid nor a fixed basis expansion of
the wave function.

\section{Theory}

\subsection{Complex gaussians}

We first consider the general case
of a quantum system composed of $N$ charged particles (e.g., a non-Born--Oppenheimer description of
a molecule with nuclei and electrons moving in a central potential \cite{Bubin2013})
whose Cartesian coordinates are collected in a $3N$-dimensional 
real vector $\mathbf{r}$. An LCG($K$) wave-function \emph{Ansatz} (prior to spin and permutation-symmetry
adaptation) at time $t$ is expressed in terms of $K$ $N$-particle complex gaussians in the following way:
\begin{equation}\label{eq:gaussian_multi}
    \psi(\mathbf{r}, t) = \sum_{k=1}^K c_k(t)
    \exp \left[-(\mathbf{r} - \mathbf{s}_k(t))^\prime \mathbf{C}_k(t)
    (\mathbf{r} - \mathbf{s}_k(t)) \right].
\end{equation}
Here, $c_k$ is a
complex linear coefficient,  $\mathbf{s}_k$ is a complex shift vector, 
and $\mathbf{C}_k$ is a $3N\times3N$ 
matrix of complex parameters. Note that all parameters, nonlinear as well as linear, are time-dependent.
Here and in the following, the prime
denotes vector and matrix transposition (without complex conjugation).
Both the real and imaginary
parts of $\mathbf{C}_k = \mathbf{a}_k + \imag\mathbf{b}_k$
are real symmetric $3N \times 3N$ matrices. The real part must be
constrained to be symmetric positive definite to ensure square-integrability.
An alternative (but equivalent) functional form of the gaussians in Eq.~\eqref{eq:gaussian_multi} is 
\begin{equation}\label{eq:gaussian_real}
    g_k(\mathbf{r}) =
    \exp \left[-\frac12 (\mathbf{r}-\mathbf{q}_k)^\prime \mathbf{C}_k
    (\mathbf{r}-\mathbf{q}_k) + \imag\mathbf{p}_k^\prime(\mathbf{r}-\mathbf{q}_k)\right],
\end{equation}
with real vectors $\mathbf{q}_k$ and $\mathbf{p}_k$ defining the center and 
momentum of the gaussians, respectively.

Both the linear expansion coefficients \emph{and} the nonlinear parameters 
of each gaussian are free variables that can be optimized in the calculation. 
In Eqs.~\eqref{eq:gaussian_multi} and \eqref{eq:gaussian_real} the exponent is a general second-order polynomial with 
respect to the complex parameters, up to an irrelevant constant.
As is well known, the family of 
gaussian functions is complete in more than one sense~\cite{Bachmayr2014_NM,Lasser2020_ANa}. 
Therefore, choosing even a modest $K$ value with freely adjustable nonlinear parameters
should form an inordinately flexible basis set to represent 
almost any wave function.

Importantly, integrals for the matrix elements of the kinetic and potential energies
and the overlap have explicit 
analytic formulas that can be straightforwardly implemented 
in an efficient computer code. For example, unshifted complex gaussians satisfy
\[
    \braket{g_k\vert r_{ij}^{-1} \vert g_l} =
    \frac{2\braket{g_k\vert g_l}}{\sqrt{\pi\,\mathrm{tr}((\mathbf{C}_k
                                + \mathbf{C}_l)^{-1}\mathbf{J}_{ij})}},
\]
where $r_{ij} = \vert \mathbf{r}_i-\mathbf{r}_j \vert$ and $\mathbf{J}_{ij}$ is the matrix associated with the quadratic form 
$r_{ij}^2 = \mathbf{r}^\prime \mathbf{J}_{ij}\mathbf{r}$ \cite{Bubin2006_TJoCP,Bubin2008}. 
Such analytic formulas pave the way for 
developing a completely grid-free gaussian time-propagation approach. 

\subsection{The Rothe method}

The time-dependent Schrödinger equation can be phrased in a variational form as
\begin{equation}
    \dot{\psi}(t) = \argmin_\chi \| \imag\chi - \hat{H}(t)\psi(t) \|, \label{eq:tdvp}
\end{equation}
where $\chi$ ranges over all allowed infinitesimal variations of $\psi(t)$.
For some approximate \emph{Ansatz} depending smoothly on a set of (real) parameters,
the variations are restricted, leading to an implicit system
of ODEs for the parameters. This is the first step 
of the Dirac--Frenkel variational principle.
In the next step, the ODEs are
integrated using some numerical integration scheme. 
It should be noted, however, that time-propagation with the
Dirac--Frenkel principle is infamous for its numerical challenges \cite{Kay1989},
which are particularly pronounced for LCG \emph{Ans\"{a}tze}, see for example  \cite{Sawada1985,Rowan2020_PRE}.
While successful work-arounds have emerged, including the frozen gaussian approximation \cite{RPSWBL2015}
and sophisticated basis re-expansion techniques  \cite{Kong2016,Saller2017,Lee2018,Tatsuhiro2018,Joubert-Doriol2022,Dutra2022},
no general approach that can fully exploit the high flexibility of complex gaussians to simulate
complicated quantum dynamics has been formulated.

Somewhat paradoxically, it is the high flexibility of the complex gaussians
that causes the numerical challenges.
There may be several distinct LCGs that 
approximate the same wave function to a similar accuracy---i.e.,
there is no unique set of the gaussians and, thus, no unique set of their nonlinear
parameters. This leads to the system
of ODEs being insoluble due to singular or severely ill-conditioned Gramian matrices, which present a challenge that is significantly
harder to overcome than ordinary ``stiffness''.

Rothe's method offers a different approach  
than the Dirac--Frenkel principle. Placing all focus on the \emph{total time-evolving
wave function}, Rothe's method de-emphasizes the evolution of 
the nonlinear gaussian parameters and the linear expansion coefficients.
Equation~\eqref{eq:tdvp} is discretized \emph{first} 
in the time variable by some suitable scheme. 
For example, using the \emph{implicit trapezoidal rule} and time step~$h$,
the wave function at time $t_n = nh$ is:
\begin{equation}
    \psi^{n} = \argmin_\varphi \| \mathcal A_n\varphi - \mathcal A_{n-1}^\dagger\psi^{n-1}\|, \label{eq:rothe}
\end{equation}
where $\mathcal A_n = I + \imag h \hat{H}(t_{n})/2$ and 
$\psi^n \approx \psi(t_n)$ is an approximation to the wave function at time $t_n$.
This turns the propagation into a
sequence of nonlinear optimization problems.
Although any other scheme can be chosen instead,
the trapezoidal rule is a rather natural starting point, see also \cite{Lopez2022}.
In particular, the exact solution to the time-propagation equation \eqref{eq:rothe} 
in the Hilbert space is given by
the Crank-Nicolson (CN) scheme~\cite{Varga2011_}, whose local error is $O(h^3)$.

Using the time-discretized version of the time-dependent 
variational principle in Eq.~\eqref{eq:rothe}, one can now introduce an \emph{Ansatz}, e.g., the LCG($K$) wave function, at each time $t_n$, $\psi^n = \psi(\alpha^n,c^n)$.
Here, $\alpha^n$ and $c^n$ denote the sets of nonlinear and linear parameters, respectively, at time $t_n$.
Some restrictions need to be imposed on the wave-function optimization problem 
to control the error in the calculation. A tolerance 
$\epsilon>0$ is chosen and used in the propagation of the LCG wave
function, $\psi(\alpha^n,c^n)$: 
\begin{equation}
    \frac{1}{2}\| \mathcal A_n\psi(\alpha^n,c^n) - \mathcal A_{n-1}^\dag\psi(\alpha^{n-1},c^{n-1}) \|^2 < \epsilon. \label{eq:objective0}
\end{equation}
The completeness of the gaussians ensures that the tolerance $\epsilon$ can always be achieved
with an adaptive number of gaussians in the basis set.
Hence, our propagation scheme for the LCG ansatz from time $t_{n-1}$ to $t_n$
needs to include a procedure for augmenting the gaussian basis set
with additional functions.

The problem~\eqref{eq:objective0} is separable 
in the sense that freezing $\alpha^n$ results in a 
\emph{linear} least-squares problem for $c^n$, see \cite{Golub1973_SJNA}. 
This leads to a reduced problem defined only in terms of
finding a set of nonlinear parameters $\alpha^n$ such that
\begin{equation}\label{eq:nonlinear_ls}
    F(\alpha^n) = \frac{1}{2}\| (I - P_{\mathcal A_{n}}) \mathcal A_{n-1}^\dag\psi(\alpha^{n-1},c^{n-1}) \|^2 < \epsilon.
\end{equation}
Here, $P_{\mathcal A_n} = P_{\mathcal A_n}(\alpha^n)$ is an orthogonal projector
on the space spanned by the gaussians 
transformed with the $\mathcal A_n$ operator, and thus an explicit function of the unknowns $\alpha^n$.
Elimination of $c^n$ is essential, as dependent variables in a nonlinear least-squares problem often 
lead to ill-conditioning of the problem~\cite{Golub1973_SJNA}.

Our nonlinear optimization scheme is a variant of 
the iterative Gauss--Newton method with step-size control via 
a simple Armijo backtracking strategy to ensure sufficient 
decrease of the objective function in each 
iteration~\cite{Nocedal2006_,Floudas2009_}. The Gauss--Newton 
method requires solution of a linear system in each 
iteration, and approaches quadratic convergence when the 
objective function becomes small. A successful use of the 
Newton method for solving the equation $\nabla F(\alpha^n) = 0$ 
relies on providing a sufficiently good initial guess
at the start of the calculation. 
In our approach, we reuse the optimized 
nonlinear parameters $\alpha^{n-1}$ 
obtained for $t_{n-1}$ as the initial guess
for the parameters at $t_n$.
This usually works quite well and improves when the time step is lowered. 
However, the nonlinear optimization routine may get trapped 
in a local minimum with $F(\alpha^n) \geq \epsilon$. 
In such a case, we add one or more gaussians to the basis set. 
Our chosen strategy 
for the basis-set enlargement is fairly simple, but works quite well  
in the test simulation reported below. The nonlinear parameters of
a gaussian are completely determined by the expectation 
values of position, momentum, position variance, and momentum 
variance. 
The residual expression, $f(\mathbf{r})$ (i.e., the function that appears 
inside the norm bars of 
$F(\alpha^n)$), provides the function 
for which the above-mentioned expectation values are calculated.
Next, a gaussian is generated whose expectation values match
the expectation values determined for $f(\mathbf{r})$.
That gaussian is added to the LCG basis set and a complete
optimization is performed for all gussians in the enlarged basis set.
If still $F(\alpha^n) \geq \epsilon$, a second gaussian is 
added to the basis set using the same procedure as used for 
adding the first 
gaussian. This addition process 
continues until $F(\alpha^n) \leq \epsilon$.
The calculation then proceeds to the next time step.

\section{Numerical results}

We use a simple one-dimensional (1D) model system to perform a proof-of-principle
test simulation, which is simple enough to allow detailed analysis of the results
while challenging enough to be impossible to carry out with conventional, straightforward
techniques based on the Dirac--Frenkel variational principle.

\subsection{1D model system}

\begin{figure}
    \centering
    \includegraphics{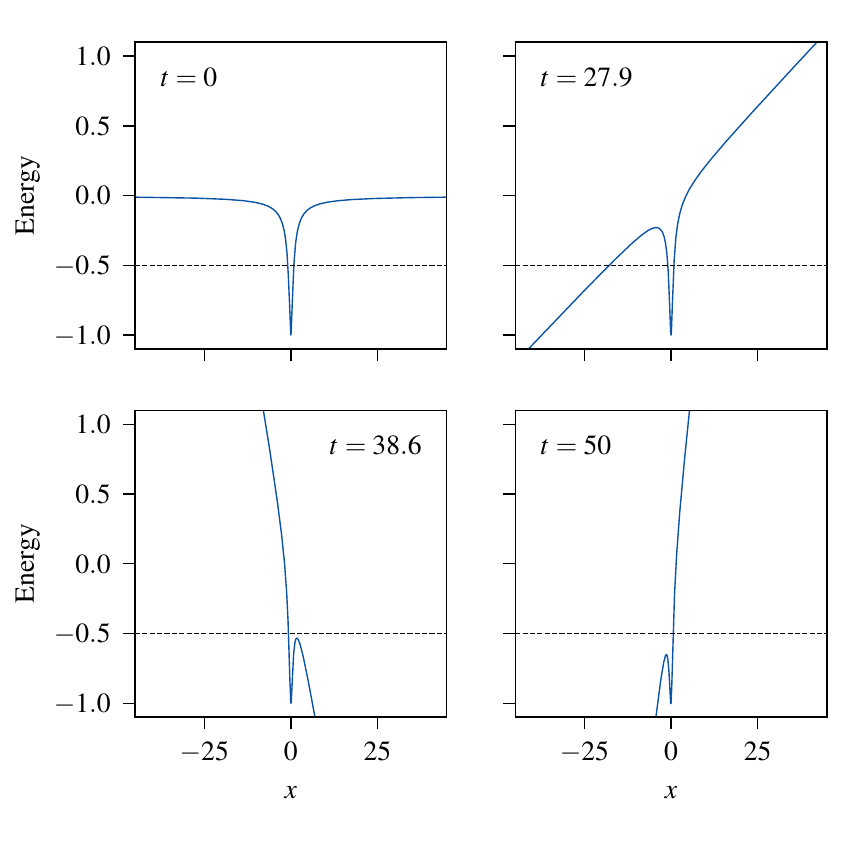}
    \caption{The effective potential $V(x) + x\mathcal{E}(t)$ at the extrema
    of the electric field during the first half of the pulse. The horizontal
    dashed line indicates the ground-state energy of the potential $V(x)$.
    \label{fig:potentials}}
\end{figure}

We consider the simplest possible model system for a proof-of-principle 
study of the Rothe method applied to an LCG \emph{Ansatz}: a one-dimensional, one-particle system.
The Hamiltonian is defined by
\begin{equation}
    \hat{H}(t) = -\frac{1}{2}\frac{\partial^2}{\partial x^2} + V(x) + x\mathcal{E}(t),
\end{equation}
where $V(x) = -(1/2)/\sqrt{x^2 + 1/4}$  mimics the nuclear potential for the electron in a hydrogen atom.
The ground-state energy in this potential is conveniently located at $E_0 = -1/2$.
Note that atomic units are used throughout except where explicitly indicated otherwise.
The time-dependent electric field $\mathcal{E}(t)$ is nonzero only 
in the $t_0 < t < t_1$ region, where it becomes equal to:
\begin{equation}
    \mathcal{E}(t) = \mathcal{E}_0\sin^2\left(\pi \frac{t-t_0}{t_1 - t_0}\right)\cos(\omega (t-\bar{t})), \quad \bar{t} = \frac{t_0 + t_1}{2}.
\end{equation}
In our model, we set $\omega = 0.25$ ($6.8\,\text{eV}$, $182\,\text{nm}$),  $t_0 = 20$, and $t_1 = 80$
(foot-to-foot duration $1.45\,\text{fs}$).
The maximum amplitude of the field is $\mathcal{E}_0 = 0.225$ ($116\,\text{V/nm}$)
and occurs at $t = 50$, corresponding to peak intensity $18\times 10^{14}\,\text{W/cm}^2$
and ponderomotive energy $0.203$. The Keldysh parameter is $\gamma = 1.11$, which is traditionally interpreted
as indicating predominance of multiphoton ionization over tunnel ionization.
We stress that
the laser-pulse parameters are chosen to generate complicated quantum dynamics, \emph{not} to emulate a particular experimental setup.
As is evident from the effective potential plotted at the extrema of the electric field in Fig.~\ref{fig:potentials},
the laser pulse violently disrupts the potential and certainly will induce significant ionization probabilities.

A highly accurate grid reference calculation is performed by spatially discretizing 
the real axis using a grid with $n_\text{grid}=4096$ 
equidistant points in the interval $[-l,l)=[-500,500)$. The kinetic-energy part is approximated 
using the standard Fast Fourier Transform (FFT) approach. 
This approach introduces artificial periodic boundary conditions, 
which have negligible effect on the results due to the large domain used in
the calculations. The time evolution
can be carried out in a number of ways. 
In the present work we choose the well-known 
Crank--Nicolson (CN) scheme~\cite{Varga2011_} with the time step $h = 10^{-3}$, 
since our proposed LCG propagation scheme is an approximation
to that scheme.

\begin{figure}
    \centering
    \includegraphics{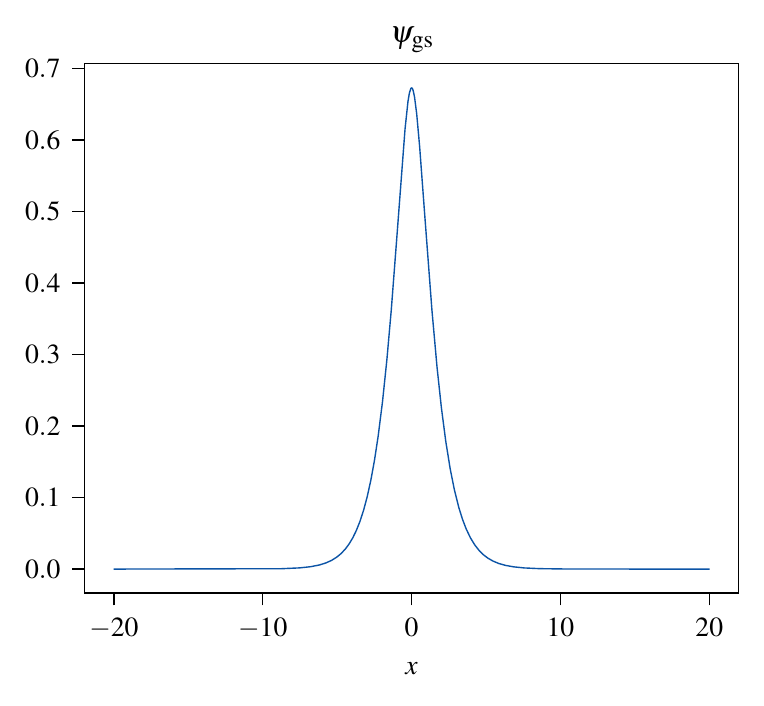}
    \includegraphics{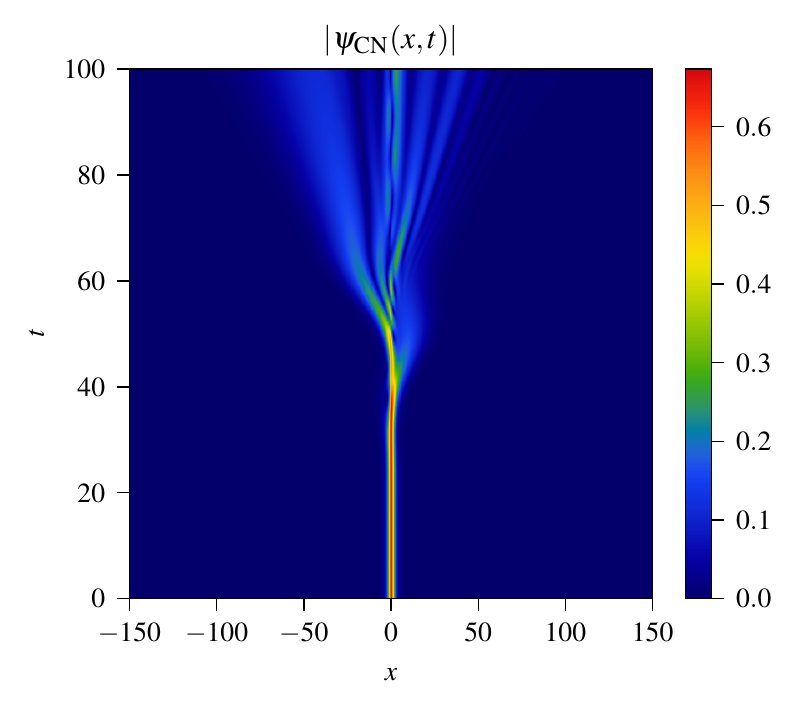}
    \includegraphics{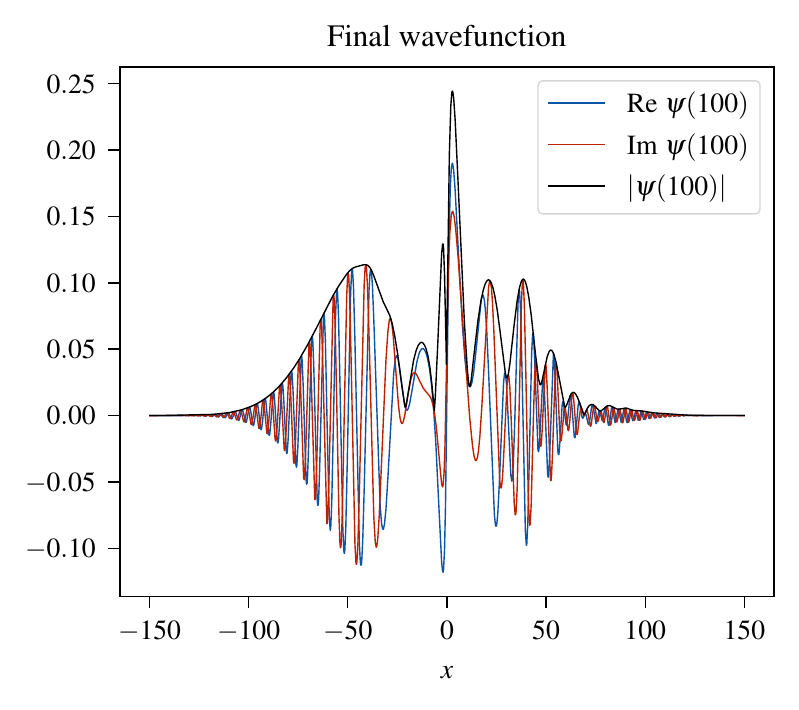}
    \caption{The initial wave function (top), the propagation history of the model system (middle), and 
    the final wave function (bottom). The propagation is performed on a fine 
    spatial grid using the Crank--Nicolson method with time step $h=10^{-3}$.}
    \label{fig:reference-calc}
\end{figure}

The initial wave function is the ground state of the model potential $V(x)$ obtained by inverse iterations 
with the conjugate gradient method \cite{Varga2011_}.
It is shown in Fig.~\ref{fig:reference-calc} along with the complete history of the propagation 
of the essentially exact grid-based wave function $\psi_\text{CN}(x,t)$ and the final 
wave function, $\psi_\text{CN}(x,100)$. As one can see, the final wave function
spreads over more than $200\,\text{bohrs}$ with highly oscillatory
real and imaginary parts associated with ionization. As can be seen in the propagation history in the middle panel of
Fig.~\ref{fig:reference-calc}, the spreading of the wave function
proceeds in accordance with the effective-potential oscillations depicted in Fig.~\ref{fig:potentials}.
As expected on physical grounds, the wave function continues to spread after the laser is switched off at $t = 80$.

\subsection{Rothe propagation of an LCG \emph{Ansatz}}

\begin{figure}
    \centering
    \includegraphics{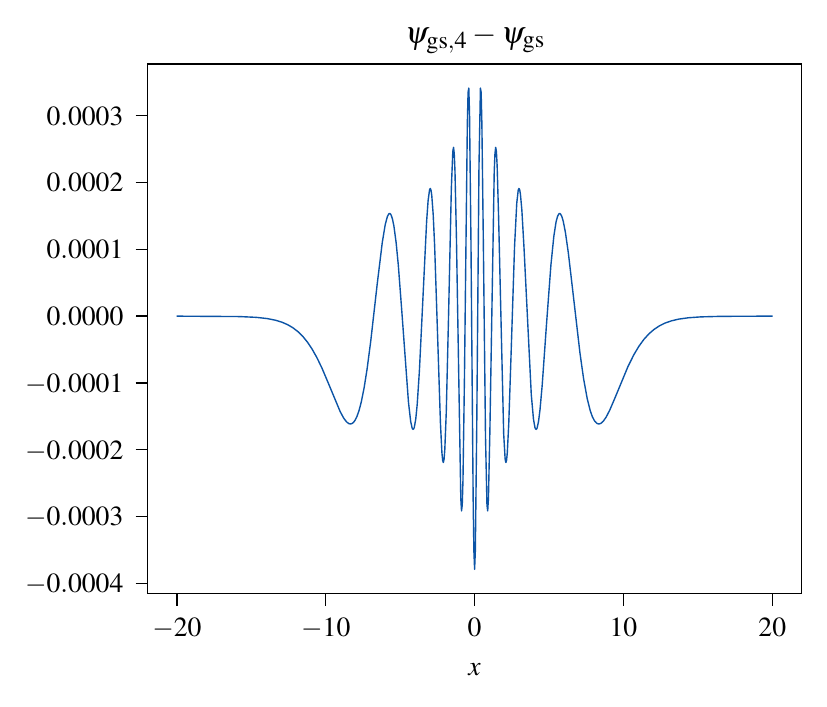}
    \caption{The local 
    error of $\psi_{\text{gs},4}$, which is an 
    LCG($4$) least-squares fit to the grid-based ground-state wave function shown
    in Fig.~\ref{fig:reference-calc}.}
    \label{fig:ground-state}
\end{figure}

In order to compare the proposed LCG propagation scheme to the
grid-based Crank-Nicolson simulation above, we need to use (very nearly) identical
initial wave functions. Therefore, we compute the LCG ground-state wave function
by a nonlinear least-squares fit to the grid-based initial state
using an LCG($4$) \emph{Ansatz}, $\psi_{\text{gs},4}$,
on the form \eqref{eq:gaussian_real} specialized to one spatial dimension.
Figure \ref{fig:ground-state} shows the resulting local error of $\psi_{\text{gs},4}$.
The real parameters of the four gaussians are: $q_i=p_i=b_i=0$ and $a = [0.37745,  
2.0681,  0.61766, 1.0688]$, and the linear coefficients are: $c = [0.08719, 0.061077, 0.29305, 0.23122]$. 
The fitting error is $\|\psi_\text{gs} - \psi_{\text{gs},4}\|^2 = 4.3485\times 10^{-7}$.
The LCG($4$) ground-state energy thus is accurate to around 7 digits.

We use the same time step as in the grid-based CN simulation, $h=10^{-3}$, and select the threshold
$\epsilon = 10^{-7}$ for the nonlinear least-squares optimization in the Rothe LCG simulation.
The value of the threshold is chosen so that the global 
error in the scheme is comparable to that used in 
the grid-based CN simulation.
In the present implementation the gaussian 
matrix elements are evaluated using quadrature
that is sufficient (and nearly exact) for the 1D model considered 
in the present calculations.

\begin{figure}
    \centering
    \includegraphics{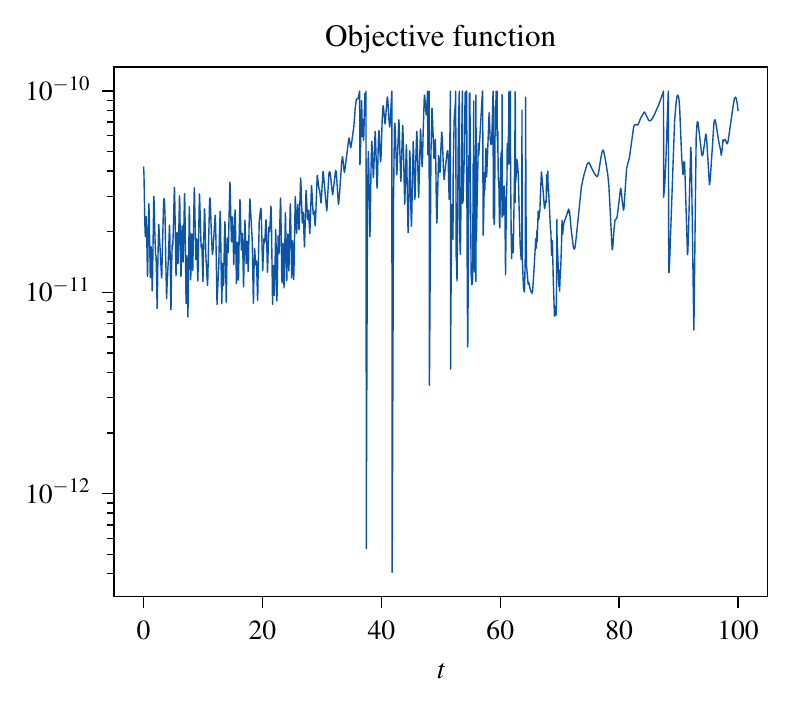}
    \includegraphics{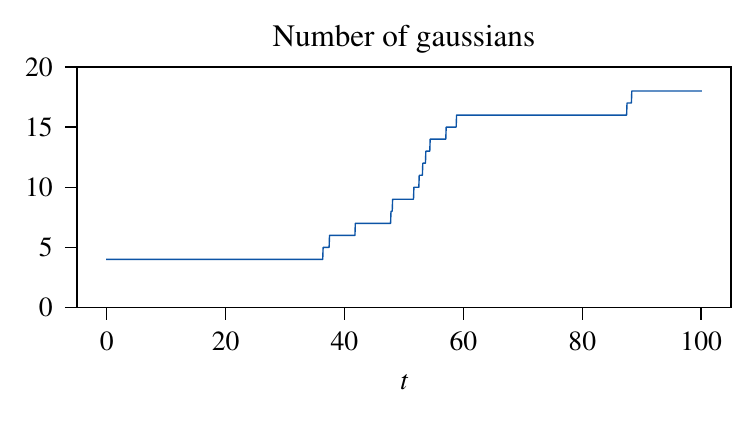}
    \caption{Objective function and number of gaussians as a function of 
    time during the LCG propagation using Rothe's method.}
    \label{fig:objective}
\end{figure}

The Gauss--Newton iterative method involves the numerical solution of linear systems of comparable size and structure as those required for evaluating the parameter time-derivatives in the Dirac--Frenkel principle. As the calculation progresses, the Gauss--Newton method 
uses a single iteration in the vast majority of the time steps 
(it uses two iterations in a tiny fraction of the time steps, 
and three and nine iterations in only two time steps). In contrast to the Dirac--Frenkel principle, the conditioning of the linear systems does not seem to be a problem.
The time propagation produces near-smooth paths for 
all gaussians except where functions are added. 
At those points, the reoptimization of all 
the non-linear parameters of all gaussians requires 
varying numbers of iterations, but they always converge without a problem. 
In Fig.~\ref{fig:objective}, the objective function $F(\alpha^n)$ 
is shown, together with the number of gaussians needed 
to achieve the convergence criterion, $F(\alpha^n) < \epsilon$,
as a function of time.  The number of functions increases 
throughout the simulation with a modest value of $K=18$ 
at the final time $t=100$. 

\begin{figure}
    \centering
    \includegraphics{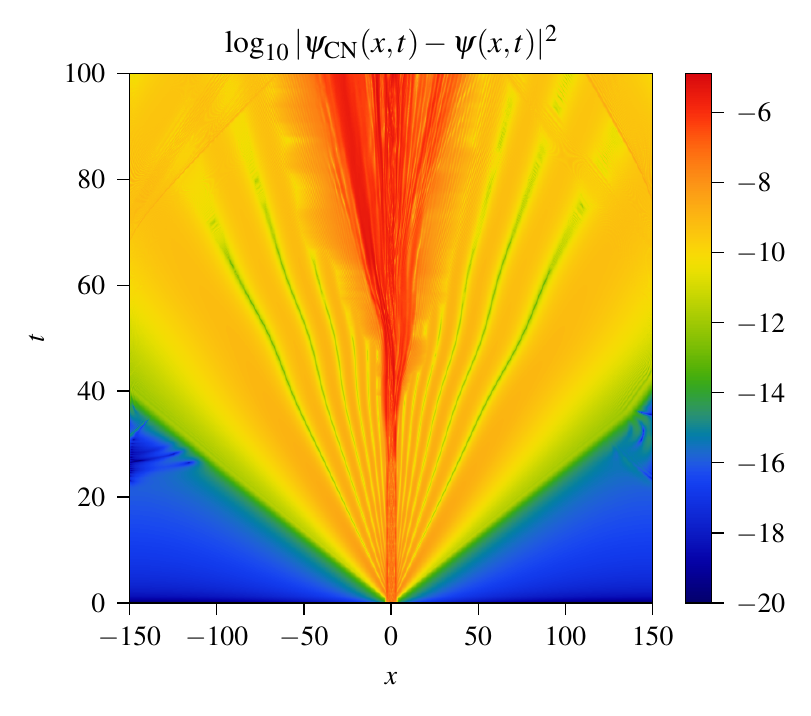}
    \caption{Local errors of the LCG propagation relative to the Crank--Nicolson propagation. Boundary artifacts from the periodic boundary conditions can be seen, but these have extremely small values.}
    \label{fig:hist-comp}
\end{figure}

\begin{figure}
    \centering
    \includegraphics{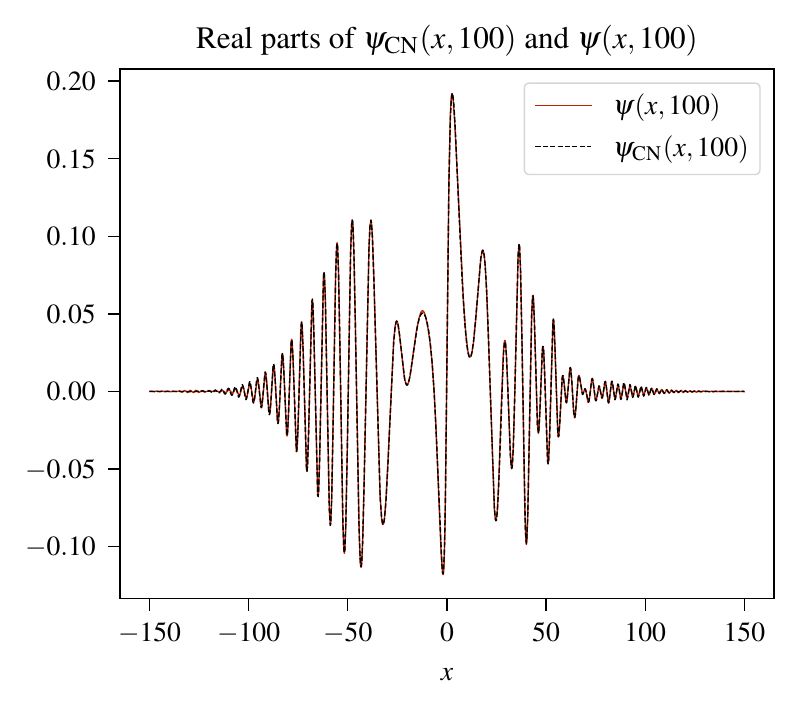}
    \includegraphics{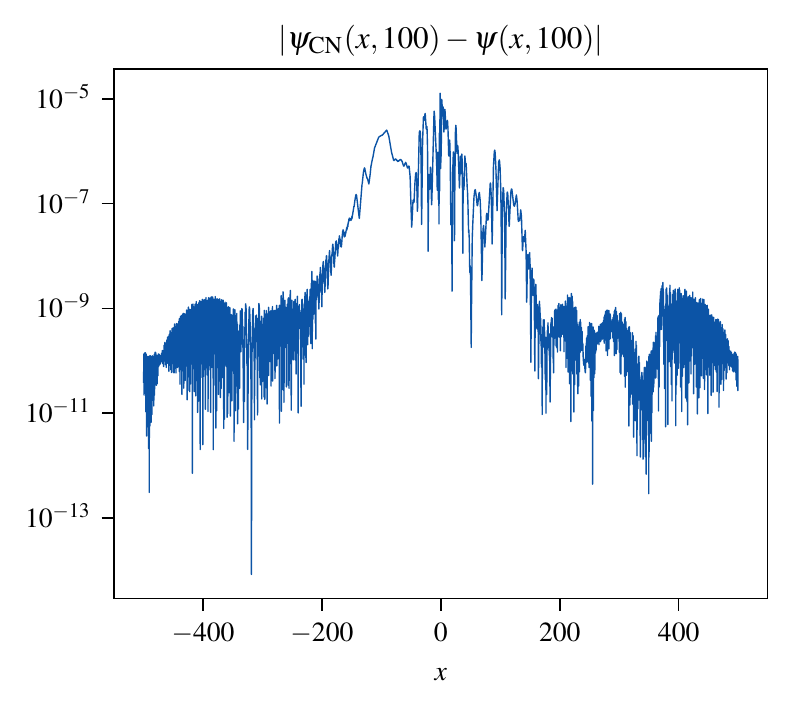}
    \caption{Top: Real parts of final wave functions. 
    At the resolution of the plot, 
    no difference can be seen. Bottom: absolute value squared of the error in the final wave function.}
    \label{fig:final-comparison}
\end{figure}

The local error of the LCG propagation relative to the 
grid-based CN propagation is shown in Fig.~\ref{fig:hist-comp}. 
The density plot corresponding to the middle panel of Fig.~\ref{fig:reference-calc} 
is not shown, as the local errors are too small to make 
the plots visibly different. The log-scale plot of the local 
error, Fig.~\ref{fig:final-comparison}, reveals that errors are present, but, in general, 
they have very small values.

The final LCG and CN wave functions are compared in Fig.~\ref{fig:final-comparison}, which reveals that the largest errors
occur around $x=0$. Still, the largest errors are an order of magnitude lower than the errors in the ground-state wave function,
Fig.~\ref{fig:ground-state}.

For each time step, on average, the most time consuming
part of the calculation is solving a set of linear equations with the
dimension equal to the number of nonlinear wave-function parameters.
In the present simulations, the number of gaussians
needed to meet the error tolerance ranges between $K=4$ and $K=18$.
Hence, the linear systems to be solved are small, and a very accurate representation of the
highly oscillatory and delocalizing dynamics of the 1D model system
is obtained.
We note in passing that for three-dimensional systems of (sets of) identical particles,
where the correct spin and permutation symmetries must be treated explicitly, the computational
bottleneck of each time step will likely be the evaluation of integrals over the gaussians rather than
solving the least-squares problem \eqref{eq:nonlinear_ls}.

The performance of the present method
compares favorably with state-of-the-art methods for the ab initio 
simulation of small atoms and molecules in terms
of efficiency and error control. For example, methods employing
a fixed basis of B-splines~\cite{Bachau2001} and/or spherical harmonics
have to either grid the computational box for the B-splines or to
set a maximal degree for the spherical harmonics.
In both cases the basis set stays fixed and has to be chosen beforehand---see,
for example, \cite{Sopena2022} for a recent simulation of
the hydrogen molecule exposed to an ultrashort high-intensity laser pulse.
Another example is the recent work of the present authors \cite{Adamowicz2022}
where a fixed basis set of real, shifted explicitly-correlated gaussians was used to
simulate laser alignment of the HD molecule without the Born-Oppenheimer approximation.
Alternative approaches, like the time-dependent
R-matrix method \cite{Lysaght2009}, the time-dependent surface
flux method (t-SURFF) \cite{Tao2012}, or the exterior complex-scaling
method \cite{Foumouo2006}, alleviate the inherent limitations of fixed basis sets by a sophisticated
approximation of the dynamics in an a priori defined outer region.
However, all state-of-the-art methods lack real-time monitoring of
the numerical time-propagation error, a hallmark of Rothe's method as implemented here.

\section{Conclusion}

We have demonstrated that a grid-free propagation scheme 
for linear combinations of complex gaussians is possible using Rothe's method.
We have introduced an efficient time-integration 
method that approximates the Crank--Nicolson scheme. 
Our initial investigation shows that this approach 
is robust and, with just $18$ gaussians, complicated wave-function dynamics 
of a hydrogenic 1D model driven by an extreme laser pulse
can be simulated with very high and controllable accuracy. To our knowledge,
propagating an LCG wave-function \emph{Ansatz} with this many flexible gaussians has never been successfully
done within the framework of the Dirac--Frenkel principle.
The scheme can be straightforwardly generalized to realistic 
models of many-particle atoms and molecules using explicitly 
correlated gaussians within or without the Born-Oppenheimer approximation.
Thus, the need for a large grid
to accurately resolve local details of the time-evolving wave function, including
dynamics that involve continuous parts of the spectrum such as ionization and dissociation,
can be completely eliminated.

\section*{Acknowledgments}

The authors acknowledge the support of the Centre for Advanced Study in Oslo, Norway, which funded and hosted our CAS research project
\emph{Attosecond Quantum Dynamics Beyond the Born-Oppenheimer Approximation}
during the academic year 2021/2022.
The work was supported by the Research Council of Norway through its Centres of Excellence scheme, Project No.\ 262695.
Partial support from the National Science Foundation (grant No. 1856702)
is also acknowledged.

\end{document}